\documentclass{svjour3}                     
\smartqed  
\usepackage[normalem]{ulem}
\usepackage{graphicx}
\usepackage{color}
\usepackage{amsmath}
\usepackage{ulem}

\usepackage{color}

 \journalname{Scientometrics}
\begin{document}

\title{
Predicting results of the Research Excellence Framework using {departmental} $h$-Index%
%
}


\author{O.~Mryglod \and R.~Kenna \and Yu.~Holovatch \and B.~Berche }

\institute{O. Mryglod \at
              Institute for Condensed Matter Physics of the National Academy of Sciences of Ukraine,
              1 Svientsitskii Str., 79011 Lviv, Ukraine\\
              \email{olesya@icmp.lviv.ua}
           \and
            R. Kenna \at
              Applied Mathematics Research Centre, Coventry University,
              Coventry, CV1 5FB, England
           \and
            Yu. Holovatch \at
              Institute for Condensed Matter Physics of the National Academy of Sciences of Ukraine,
              1 Svientsitskii Str., 79011 Lviv, Ukraine
           \and
            B. Berche \at
               Universit\'e de Lorraine, Campus de Nancy, B.P. 70239, 54506 Vand\oe uvre l\`es Nancy Cedex, France
}

\date{Received: 7 November 2014 / Accepted: date}

\maketitle

\begin{abstract}
We compare estimates for past institutional research performances coming from two bibliometric indicators to the results of the UK's Research Assessment Exercise which last took place in 2008.
We demonstrate that a version of the {departmental} $h$-index is better correlated with the actual results of that peer-review exercise than a competing metric known as the normalised citation-based indicator.
We then determine {the corresponding} $h$-indices for 2008--2013, the period examined in the UK's Research Excellence Framework (REF) 2014.
We place herewith the resulting predictions on the arXiv in advance of the REF results being published (December 2014).
These may be considered as unbiased  predictions of relative performances in that exercise.
We will revisit this paper after the REF results are available and comment on the reliability or otherwise of these bibliometrics as compared with peer review.
%

%
\end{abstract}

\newpage
\section*{Introduction and motivation}
The {\emph{Research Excellence Framework}} (REF) is a peer-review based exercise in which the quality of research emanating from  universities and {higher education institutes (HEI)} in the UK  is estimated.
Such exercises take place {every four to seven years}.
%
%
and are the bases on which governmental funding is directly allocated.
They are  also the primary source for research rankings  and therefore contribute to the reputations of universities, departments and research institutes.
In fact, such exercises are by far the most important funding- and reputation-related events for UK-based research groups and their managers in the academic calender.

The REF is, however, an expensive, time-consuming and disruptive exercise, as was its previous incarnation, the {\emph{Research Assessment Exercise}} (RAE).
For this reason, suggestions have been made to replace such peer-review  systems by one based on scientometrics, or at least to include bibliometric based measures in the exercise.
Such proposals have met with stout resistance from the academic community, so far with considerable success.

The results of the next evaluation exercise are due to be announced on 18 December 2014.
Here, after comparing  bibliometric-based measures with previous RAE outcomes, we use the best of these to predict some of the outcomes of REF.

In particular, we examine bibliometric indicators on an institutional basis for four subject areas: biology, chemistry, physics and sociology.
We show that  a certain version of {the Hirsch} index \cite{Hirsch}, known as the {\emph{departmental $h$-index}} \cite{Dorothy_blog} {has a} better correlation with the results of RAE, compared with another citation-based {indicator \cite{Evidence_web}} for which  a sophisticated normalization procedure was implemented.
We then determine {departmental} $h$-indices {for  different HEI's} based upon their outputs in these subject areas in the run up to REF~2014.
We use this to rank universities in these subject areas.
Since we generate our $h$-rankings {\emph{before}} the 18 December 2014, they may be considered  an unbiased {\emph{prediction}} of the outcome of REF~2014.
Our aim is to determine whether or not the $h$-index (at least in the form used here) could have been employed as a reasonable proxy for the REF.

The preprint of paper will appear in two versions.
With the  first version, we placed our predictions on the arXiv in {\bf{November 2014}}, well after the peer reviews for REF have taken place but before the results are announced.
After 18 December 2014 we will revisit the paper and comment on the accuracy or otherwise of the $h$-prediction.

\section{Peer review versus scientometrics}

Correlations between RAE scores and different citation based metrics were studied by many different authors and comentators, including  in refs.\cite{Dorothy_blog,MacRoberts1989,Oppenheim1996,Holmes2001,2006_Raan}.
While some claimed good correlations between the resultant rankings, others point to the futility of attempts to substitute peer review by any system based on citation counting, due to identified weaknesses of citation analysis.
Recently, we also studied the correlation between the results of the most recent assessment procedure -- RAE~2008 --  and the so-called {\emph{normalized citation impact}} (NCI) \cite{2013_Scientometrics,2013_2_Scientometrics}.
The latter is a measure provided by \emph{Thomson Reuters Research Analytics} (previously known as \emph{Evidence} \cite{Evidence_web}).
We found that, for a number of disciplines, citation-based measures may inform, or serve as a  proxy for, peer-review measures of the {\emph{total strengths}} of research groups.
This means that the RAE~2008 scores scaled up to the actual size of a department correlate with the product of the NCI with  the number of staff submitted to the exercise.
The correlation is stronger in the hard sciences.
However, if  {\emph{research quality}} is defined as strength per head, we
also found that rankings based on the calculated citation impact differ significantly from the corresponding rankings based on the reported RAE~2008 scores.
In other words, while the NCI might be a reasonable indicator of departmental strength, it is not a reliable measure of relative quality (which is strength per head).

Recently, however, Bishop reported interesting results claiming relatively good correlations between the RAE~2008 quality scores for psychology and the corresponding departmental Hirsch indices  based on {\emph{Web of Science data}} for the assessment period \cite{Dorothy_blog}.
Therefore, the question of the potential suitability of citation-based metrics as a proxy for expert judgements of quality remains open.
Here we expand upon the analysis of Ref.~\cite{Dorothy_blog} for several other disciplines.
We show that the {departmental} $h$-index is indeed superior to the NCI in that it is better correlated with the results of peer review, at least for RAE~2008.
Having established the superiority of the $h$-index in this regard, we use it to predict quality rankings for the results of REF~2014.
 In doing so, we directly tackle two questions raised in Ref.~\cite{Dorothy_blog}: how well the $h$-indices {of university departments} correlate with RAE outcomes for other subjects and whether it can predict results from {submissions of research groups to} the REF.
We intend to revisit our predictions after the results of the REF become known to decide whether or not departmental $h$-indices can reasonably be used as part of, or as guidance for, the REF or as an inter-REF navigator.

\section{The RAE, the REF and citation metrics}

Since 1986, the distribution of funds for research in the United Kingdom is heavily based on the results of  peer-review based assessment procedures -- first the RAE and more recently the REF.
The results of the last RAE were published in 2008 \cite{RAE2008_web} and those of the REF will be announced  in December 2014 \cite{REF2014_web}.
At RAE~2008, academic disciplines were divided into 67 categories called units of assessment (UoA).
Higher education institutes were invited to submit researchers to any of these categories for examination by expert panels.
For REF~2014 only 37 UoA's are used.
In each assessment however, biology, chemistry, physics and sociology were included in the list of UoA's, so it is reasonable to examine these in both exercises.

For the RAE, as for the REF, the most important consideration is the quality of selected research outputs (usually in the form of published academic papers).
{RAE and REF submissions can involve research groups or centres, which are not always identical with university departments and not all members of a group have to be submitted.
Moreover, while individuals submitted to RAE/REF have to be university employees on a given census date, their submitted papers may have been published while at a previous institution, so long as the dates of publication fall inside the given RAE/REF window. }Four outputs per {submitted}  individual were {subject for} evaluation, with allowances made for part-time staff and staff with career breaks.
The RAE and REF have extensive guidelines on how to deal with matters such as collaborative research.
Publications resulting from collaborations between universities could usually be submitted by each institution.
The rules for publications involving different authors within a university depended upon whether co-authors belong to the same, or different, departments and  varied considerably across  disciplines (see also \cite{2013_2_Scientometrics}).

In addition to the quality of research outputs, the RAE sought to measure aspects of the environment and esteem associated with submitting departments and institutes. The REF is also interested in research environments but the esteem element has been replaced by measures of impact outside academia (e.g., onto industry or the public at large).
For RAE~2008 and for REF~2014, the outcome of the process was a graded profile for each submitted department or research group.
These quantify the proportion of work which falls into each of five quality bands~\cite{RAE2008_web}. The highest is denoted 4* and represents world-leading research.
The remaining bands are graded  through  3*, 2* and 1* with the lowest quality level termed ``Unclassified''~\cite{RAE_statement_panelE}.
Governmental quality-related funding is then determined by combining the profiles  in a weighted manner \cite{2013_Scientometrics}.

To determine the funding allocated to each university for their various submissions, a formula is used to convert the weighted profile into a single number $s$, which may be considered as representing a measure of overall quality of the group.
If the size of a  research group, measured by the number of submitted staff,  is denoted by $N$, its overall strength is then $S=sN$ and
the amount of quality-related funding allocated to each group is proportional to this number.
The  quality formula  was subject to regional and temporal variation post  RAE~2008.
However, immediately after the RAE~2008 results were announced, the funding formula used by the {\emph{Higher Education Funding Council for England}} (HEFCE) was  \cite{HEFCE_web}
\begin{equation}
s=p_{4*}+\frac{3}{7}p_{3*}+\frac{1}{7}p_{2*}\,,
\label{eq1_funding_formula}
\end{equation}
where $p_{n*}$ represents the percentage of a group's research which was rated $n*$.
Political pressure and lobbying resulted in a change to Eq.(\ref{eq1_funding_formula}) in an attempt to concentrate funding in the best performing universities. This resulted in usage of the alternative formula
\begin{equation}
s'=p_{4*}+{\frac{1}{3}}p_{3*}.
\label{eq2_funding_formula}
\end{equation}

Here, as in ref.~\cite{2013_Scientometrics}, we consider $s$, as defined in Eq.(\ref{eq1_funding_formula}), as a good representation of the peer-review measure of the quality of a research group. This is also the measure considered by Bishop in Ref.~\cite{Dorothy_blog}. However, in order to test the importance of the different weighting procedures, we also consider  $s'$ in what follows.
(However, since $s^\prime$ came about after political lobbying, and since it values 2* research as equal to unclassified research, we view $s^\prime$ as  a less fair and less useful measure than the original quantity $s$ -- see also Ref.~\cite{Enderby}.)

We wish to compare $s$ and $s^\prime$ to two citation-based metrics $h$ and $i$ which we explain below. Since  $h$ and $i$ are based entirely on the citations and, therefore, on research outputs (normally publications), they do not contain a direct measure of estimates of environment esteem or non-academic impact unlike $s$ and $s^\prime$.
For this reason we also use  $s_{\mathrm{output}}$ which is determined using equation~(\ref{eq1_funding_formula}) but taking into account only the output sub-profile (i.e., discarding the environment and esteem elements).


The  normalized citation impact (NCI), denoted by $i$ here, is a citation-based indicator developed by Thomson Reuters Research Analytics as a measure of departmental academic impact in a given discipline.
In refs.~\cite{2013_Scientometrics,2013_2_Scientometrics}  $i$ was compared to the results of expert assessments.
NCI values were determined for various universities using  {\emph{Web of Knowledge}} citation data \cite{Evidence_2010,Evidence_2011}.
To compare sensibly with the UK's peer-review mechanism, only the four papers per individual which were submitted to RAE~2008 were taken into account in order to determine the average NCI for various research groups \cite{Evidence_2011}.
 An advantage of the NCI is in the non-trivial normalization (so-called ``rebasing'') which takes into account the different citation patterns between different academic disciplines \cite{Evidence_2011}.
 The NCI is a relative measure (see, e.g., \cite{Schubert1996}), since it is calculated by comparing to a mean or expected citation rate.
It is also a specific measure of  academic citation impact because it is averaged over the entire research group.

Here, as in \cite{Dorothy_blog}, a departmental  $h-$index of $n$ means that $n$ papers, authored by staff from a given department, {and in a given subject area}, were cited $n$ times or more in a given time period.
{Therefore all researchers (not only those submitted to RAE or REF) publishing in a given subject area can, in principle, contribute to the departmental $h$-index.
Moreover, so long as a paper is published inside the RAE/REF window, the author address at the time of publication -- not at the REF census dated -- determines which to HEI a given output is allocated for the purpose of determination of the departmental $h$-index.} We calculate departmental Hirsch indices $h$ for groups which submitted to RAE~2008 within the selected disciplines of biology, chemistry, physics and sociology.
The citation data is taken  from the Scopus database \cite{Scopus_web}. In order to roughly calculate $h$ the following steps were performed to filter the documents:
(i) only publications which correspond to United Kingdom were considered;
(ii) to compare with RAE~2008, the publication period was limited to  2001--2007;
(iii) the following subject areas were chosen using Scopus subject categories which are closest to the RAE~2008's definition of the corresponding UoAs.
For the biological sciences, we combine the Scopus subject categories  `Biochemistry, Genetics and Molecular Biology', `Agricultural and Biological Sciences' and `Immunology and Microbiology'.
The categories `Chemistry' and `Chemical Engineering' are deemed to correspond to the RAE/REF chemistry UoA.
Similarly `Physics and Astronomy' corresponds to  the physics UoA
and `Social Sciences' to the sociology UoA.
(iv) only publications, affiliated to a particular HEI were taken into account.
{Regarding the last step, some HEI's submitted to a particular unit of assessment of RAE~2008 are sometimes absent in the Scopus `Affiliation' list. For these the values of $h$-indices can not be determined and, therefore, the numbers of HEI's in section~\ref{sectRAE2008} (Table~\ref{tab1} and Figs.~\ref{fig1_s_vs_h}--\ref{fig3_corr_dyn}) are slightly different from numbers of HEI's in section~\ref{sectREF2014} (Tables~\ref{tab2}--\ref{tab2_Soc}). To give an example, at the moment of data collecting the Scopus citation data for \emph{Open University} was available only for  papers published between 2001 and 2007, and unavailable for papers published between 2008 and 2013. Therefore, the Open University is included into the figures as well as into Table~\ref{tab1}, but excluded from the list in the Table~\ref{tab2}.}

The RAE~2008 covered research generated {in the time-window} 2001 and 2007.
We define $h_{2008}$ to be the value of the $h$-index measured at beginning of 2008.
We call this the \emph{immediate $h$-index} since it is calculated immediately after the RAE submission deadline  and only takes into account publications within the previous seven years.
{The relevance of the publication window and its effect on the  $h$-index was  discussed in Ref.~\cite{Schreiber}.}
Here we compare the metric to  $i$, $s$, $s^\prime$ and $s_{\mathrm{output}}$.
If the $h$-index were to be a useful proxy for,  or guide to,   peer review, one would require the immediate $h$-index to deliver useful information.
However, we also consider $h_{2009}$ (based on the citations made to the end of 2008), $h_{2010}$ and so on in order to determine the effects of time lags.

\section{Correlations between scientometrics and results of RAE~2008 }
\label{sectRAE2008}

We therefore have at our disposal five measures:
$s$, $s'$, $s_{\mathrm{output}}$, $i$ and $h_{\mathrm{20xx}}$, where 20xx refers to the years 2008--2014.
The first {set of} three scientometrics are based on peer review, accepted as  the ``gold standard'' in the research community.
{They apply to the research  submitted to RAE or REF.} The {second set, containing the} last two {measures,} are citation-based bibliometrics.
{They apply to research emanating from HEI's in certain Scopus-defined subject categories.
Although the research outputs are not necessarily identical with the those submitted to RAE/REF, one may reasonably expect some overlap.}
Our objective is to compare between and within the two sets.

\begin{table}[b]
\caption{The values of Pearson coefficients $r$ and Spearman's rank correlation coefficients $\rho$, calculated for different disciplines for different pairs of measures. All values are statistically significant at the level $\alpha=0.05$. {The numbers of HEI's for each discipline are given in parentheses.}}
\begin{center}
\begin{tabular}{|l|l|l|l|l|l|l|l|l|}
\hline & \multicolumn{2}{|c|}{$s$ vs. $h_{2008}$}%
&\multicolumn{2}{|c|}{$s'$ vs. $h_{2008}$}%
&\multicolumn{2}{|c|}{$s_{\mathrm{output}}$ vs. {$h_{2008}$}}%
&\multicolumn{2}{|c|}{$s$ vs. $i$}\\
\hline & $r$ & $\rho$&$r$ & $\rho$& $r$ & $\rho$& $r$ & $\rho$\\
\hline Biology {(39)}&0.7{4}&0.7{8}&0.7{4}&0.7{9}&0.6{5}&0.7{1}&0.6{7}&0.6{2}\\
\hline Chemistry {(29)}&0.{80}&0.8{3}&0.7{9}&0.8{3}&0.7{4}&0.7{1}&0.5{8}&0.5{9}\\
\hline Physics {(34)}&0.5{5}&0.5{8}&0.5{7}&0.5{9}&0.4{4}&0.4{9}&0.37&0.4{7}\\
\hline Sociology {(38)}&0.6{2}&0.5{8}&0.6{1}&0.5{8}&0.5
{7}&0.5{3}&0.5{1}&0.4{9}\\
\hline
\end{tabular} \label{tab1}
\end{center}
\end{table}

The results are summarised in Table~\ref{tab1} where Pearson's correlation coefficients and Spearman's rank correlation coefficients for various combinations are listed.
The first observation is that, although even the highest value in the table does not exceed 0.8,  $i$ is consistently less well correlated with the various peer-review-based scores than is the Hirsch index.
Since the normalization encoded in $i$ is expected to reduce the imperfections of citation counting, its poor performance is perhaps unexpected.
This surprise is compounded by the fact that $i$ determined by  taking into account the actual papers which were submitted to RAE~2008, while  the filtering of publications used to calculate $h_{2008}$ less resembles the RAE.
{The second observation is the relatively good correlations achieved by the departmental $h$-index. This is also surprising because  the Hirsch index, unlike quality measures $s$, $s^\prime$ and $s_{\mathrm{output}}$, is {\emph{a priori}} not expected to be intensive or specific.}

\begin{figure}[!t]
\centerline{\includegraphics[width=0.95\textwidth]{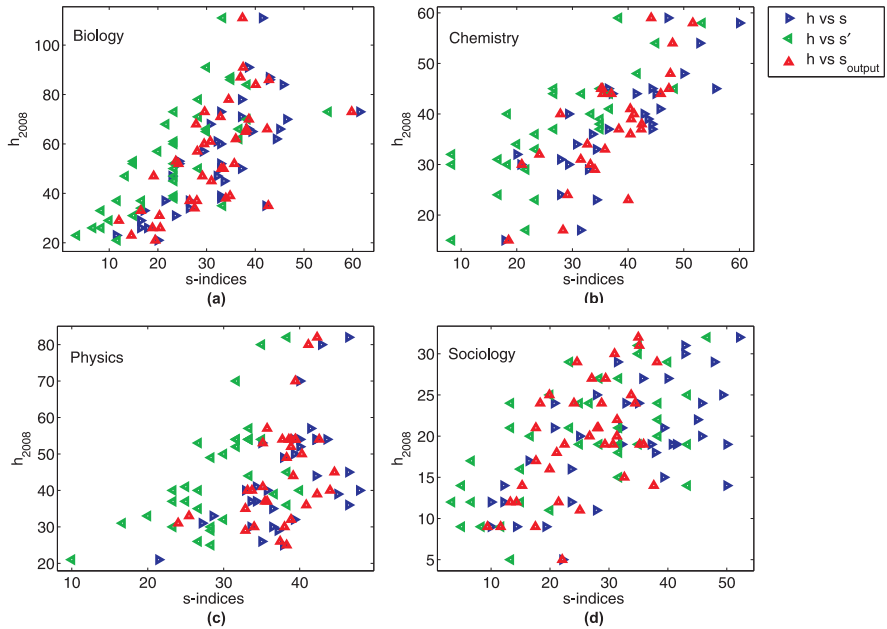}}
\caption{Correlations between {$h_{2008}$} and the peer-review based measures $s$, $s'$ $s_{\mathrm{output}}$ for research {from different HEI's in (a) biology; (b) chemistry; (c) physics and (d) sociology. }}
\label{fig1_s_vs_h}
\end{figure}

Visual representations of the correlations between the various $s$-type indices and $h_{2008}$ are given in Fig.~\ref{fig1_s_vs_h}.
We also observe that, while the relatively good correlations between the group $h$-index and the various peer-review metrics are quite similar to each other, the best match is between $s$ and  {$h_{2008}$}.
Therefore, we agree with the remark by Bishop in Ref.~\cite{Dorothy_blog} that ``the resulting $h$-index predicted the RAE results remarkably well''.

\begin{figure}[!h]
\centerline{\includegraphics[width=0.48\textwidth]{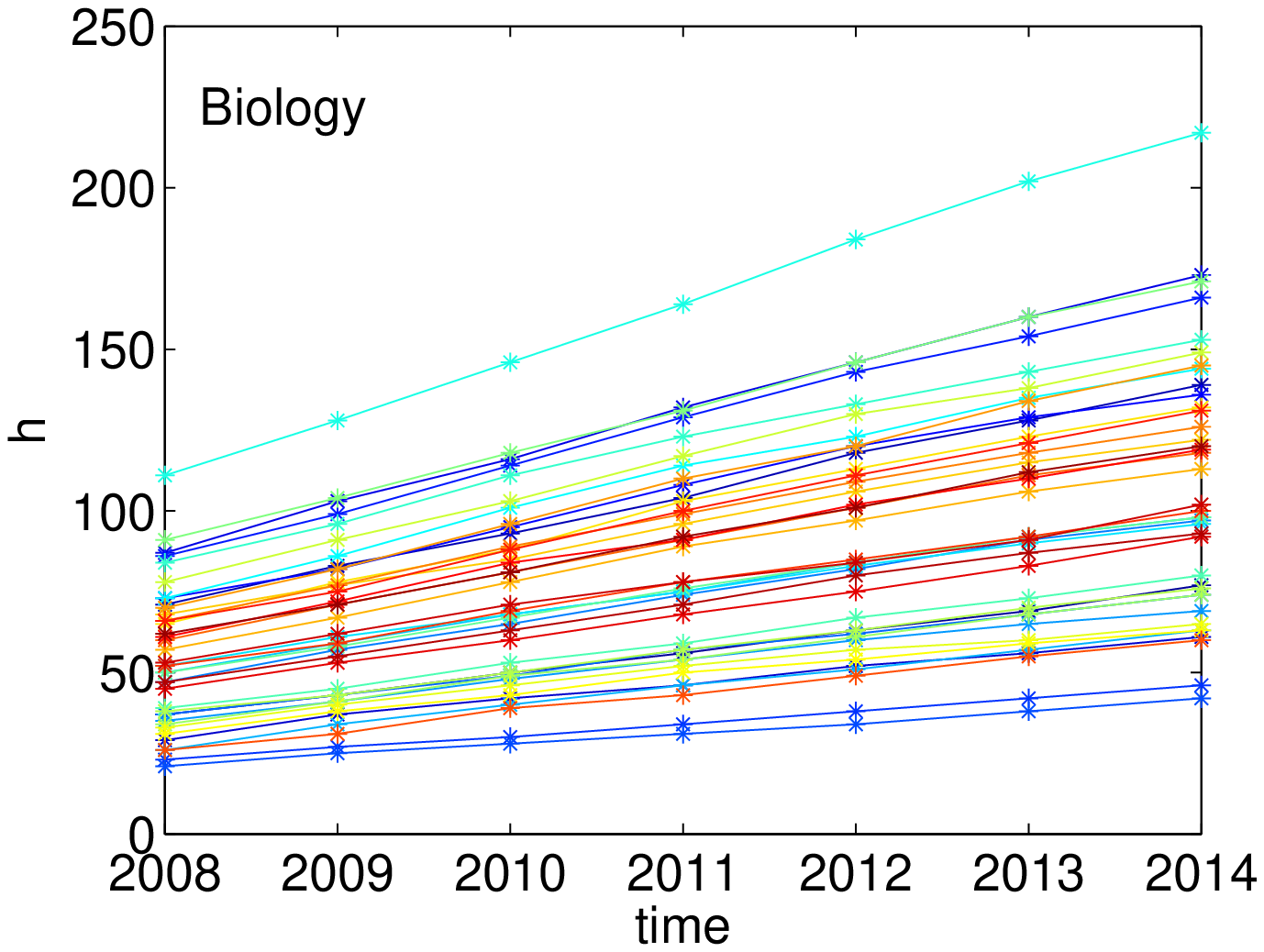}
\includegraphics[width=0.48\textwidth]{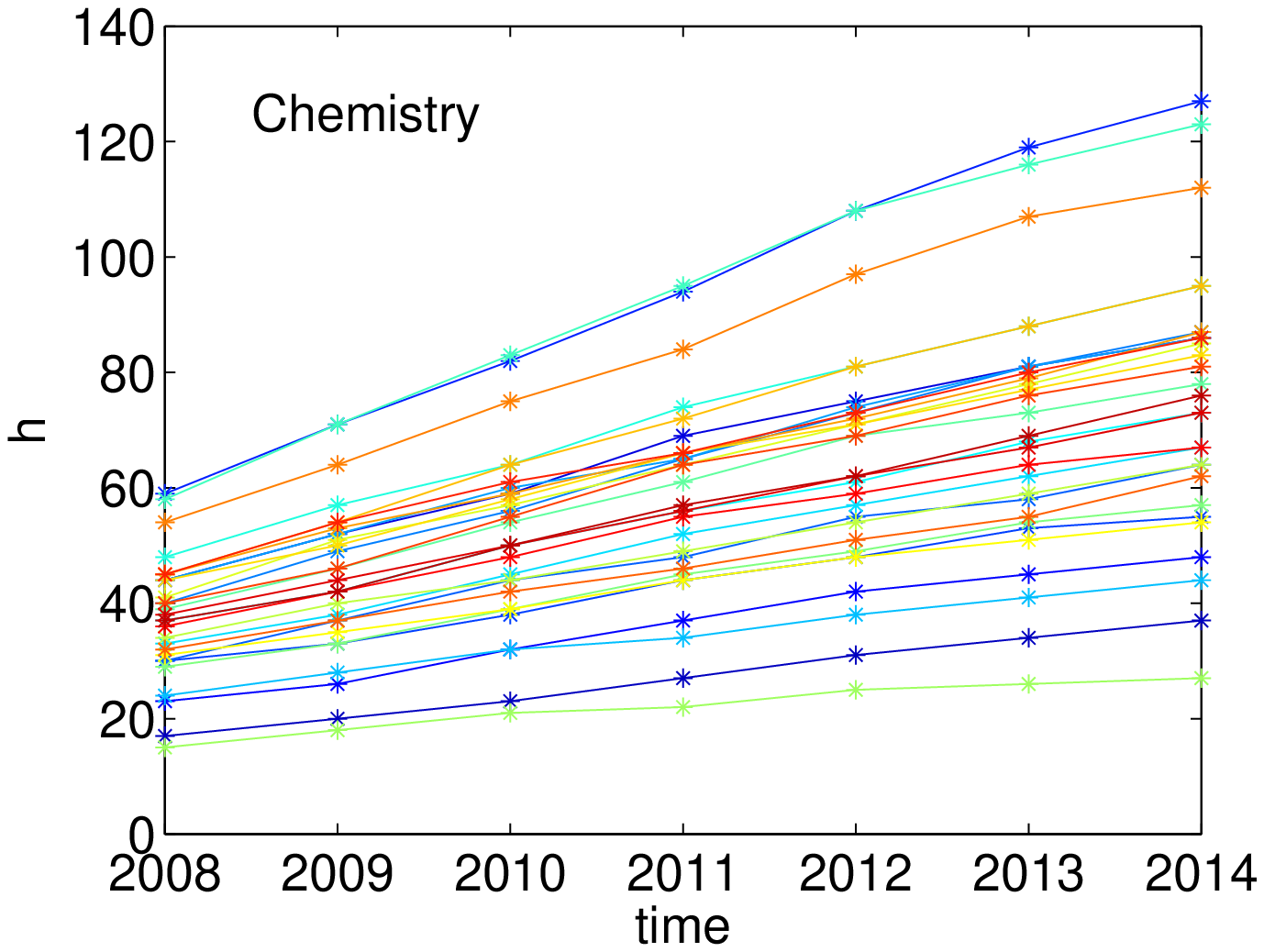}}
\centerline{(a)\hspace{4cm} (b)}
\centerline{\includegraphics[width=0.48\textwidth]{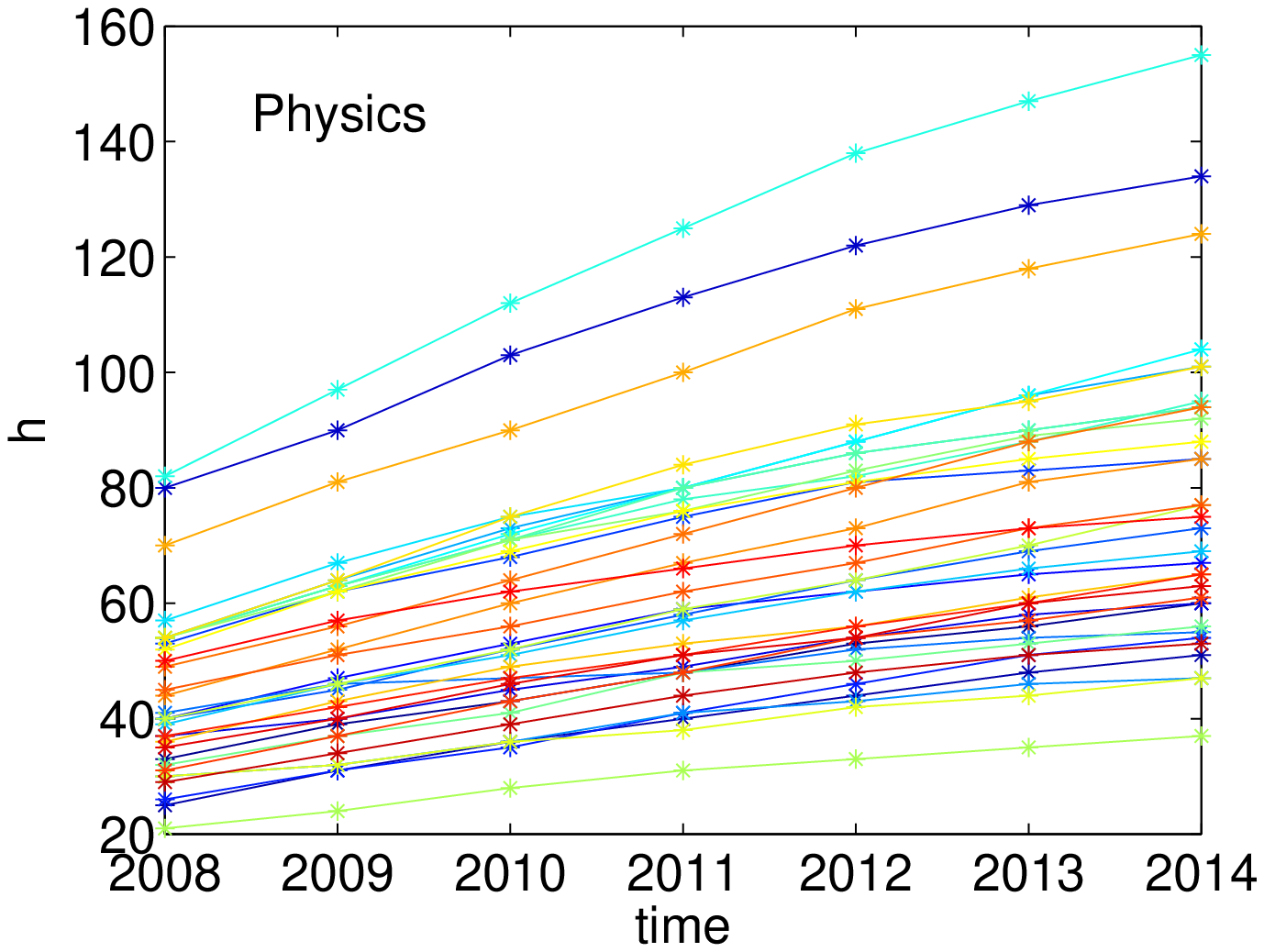}
\includegraphics[width=0.48\textwidth]{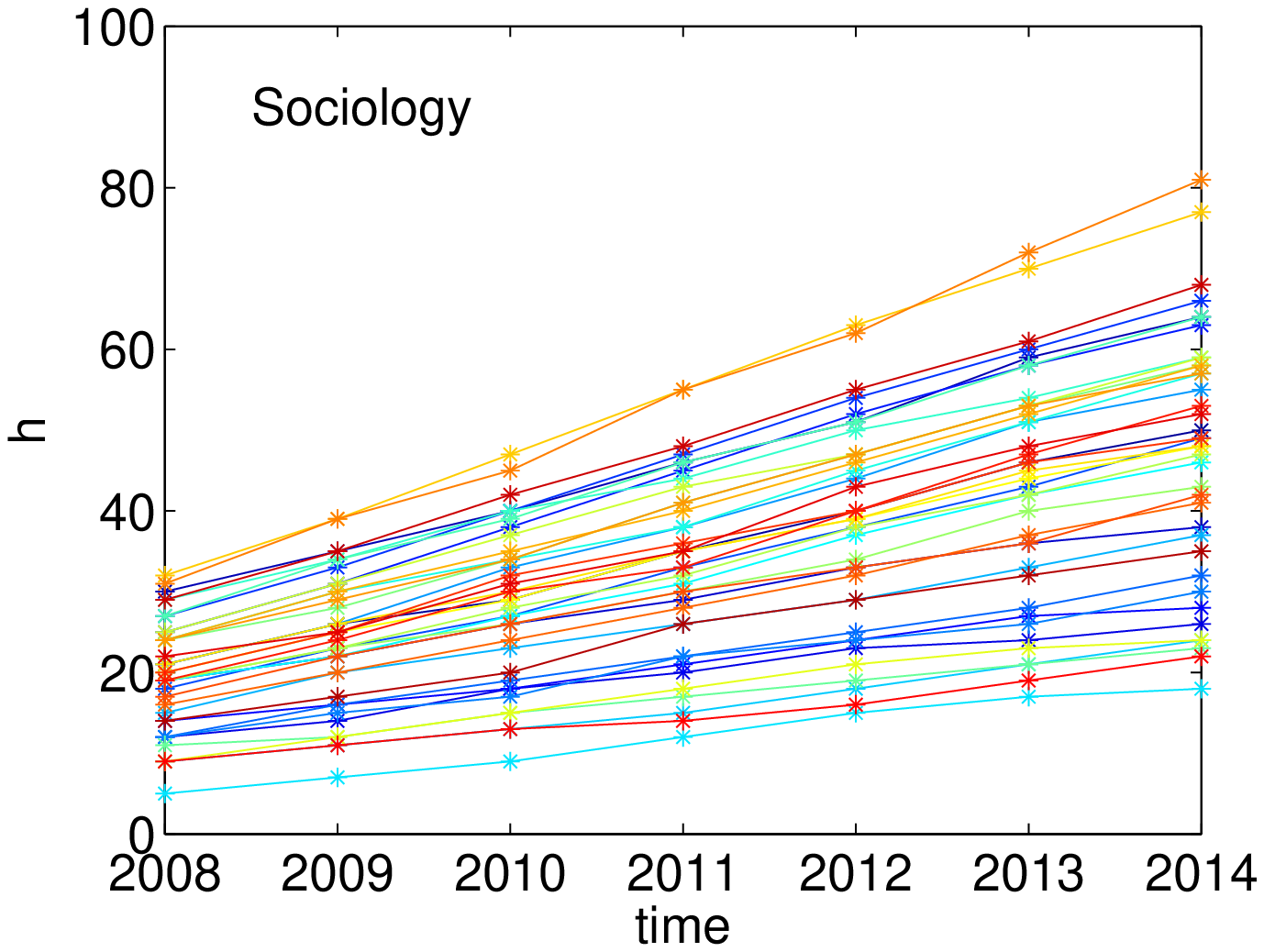}}
\centerline{(c)\hspace{4cm} (d)}
\caption{The evolution of the group $h$-indices in time ($h_{2008}$, $h_{2009}$ \dots $h_{2014}$) for (a) biology; (b) chemistry; (c) physics and (d) sociology. Different colours represent the data for different universities.  }
\label{fig2_h_dyn}
\end{figure}

A common objection to the usage of citation-based metrics is that, presumably unlike peer review, it takes a certain amount of time for citations to accumulate.
Presumably also, the time lapse is discipline dependent.
If this were to have a significant effect, one may expect increasing reliability of citation-based metrics with time elapsed since the publication window. (We are not referring about the effects of delayed recognition here since this is rather a different phenomenon.)
To check this  at the level of departments and research groups, we calculate  $h$-indices based on different time lapses: $h_{2008}$, $h_{2009}$, $h_{2010}$, $h_{2011}$, $h_{2012}$, $h_{2013}$ and $h_{2014}$.
We then plot these in Fig.~\ref{fig2_h_dyn} to track  the evolutions of departmental $h$-indices with time.
While the values of the $h$-indices grow gradually and more or less linearly with time, the ranks of the various institutions do not change significantly.

\begin{figure}[!h]
\centerline{\includegraphics[width=0.9\textwidth]{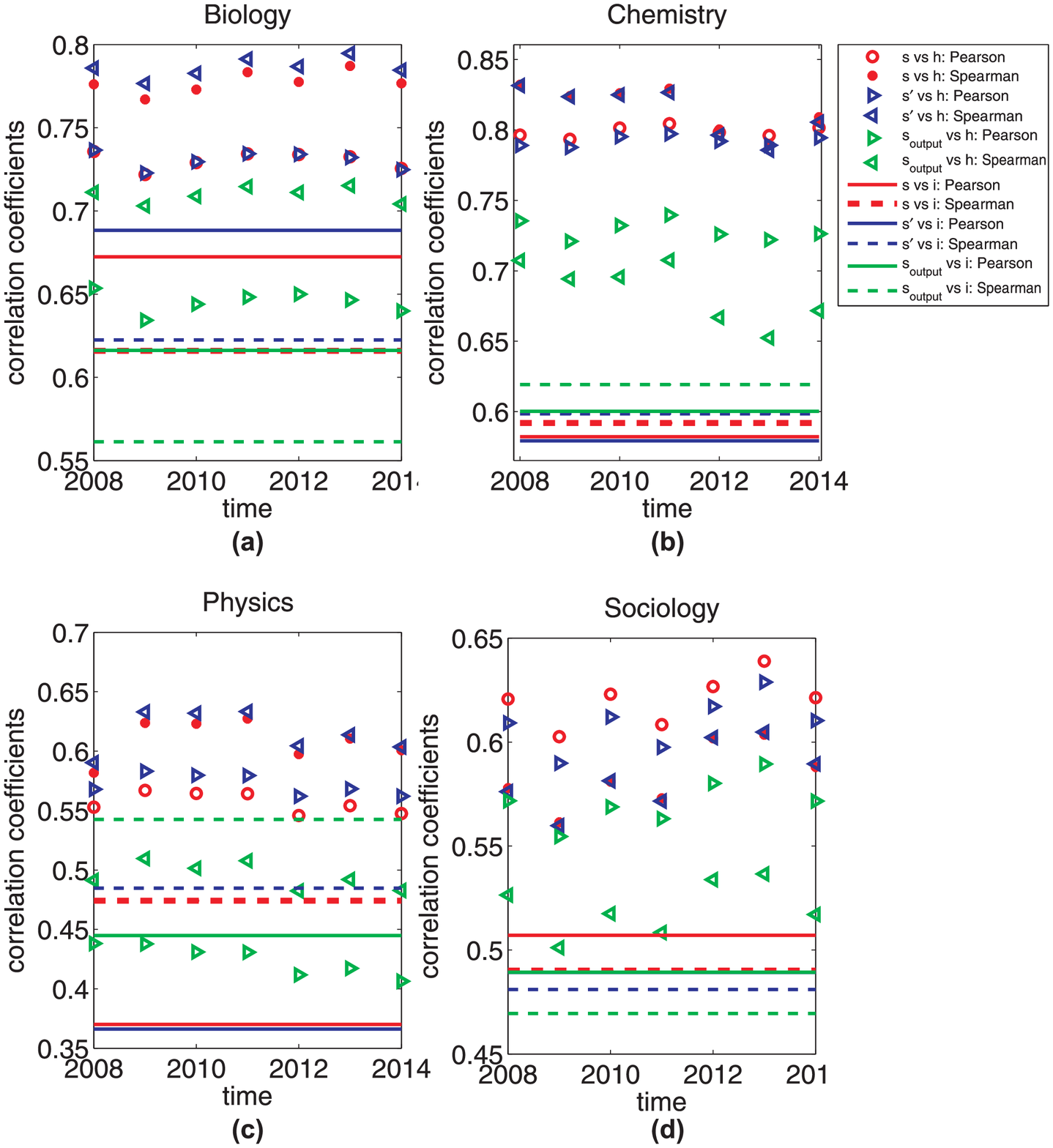}}
\caption{Pearson and Spearman correlation coefficient values between $s$, $s'$ and $s_{\mathrm{output}}$ vs $h$-indices calculated for different years {for research groups from different HEI's:} (a)
biology, (b) chemistry, (c) physics and (d) sociology. 
The corresponding values for $s$, $s'$ and $s_{\mathrm{output}}$ vs. $i$, as calculated for RAE~2008 data, are shown by lines. }
\label{fig3_corr_dyn}
\end{figure}

The dynamics of the calculated values of Pearson and Spearman coefficients are shown in Fig.~\ref{fig3_corr_dyn}. One sees that the correlations between the $h$- and $s$-values do not become noticeably stronger with time.
Moreover, the correlations between $h$ and $s$ are  consistently better than those between $i$ and $s$ for all disciplines studied.
This reinforces our earlier conclusion that RAE~2008 scores, as well as ratings built on this basis, are better correlated with departmental $h$-indices than with the normalized citations impact $i$. Moreover, and importantly, it is reasonable to use the immediate $h$-index, which can be calculated right away after the end of the fixed publication period -- one does not have to wait years for citations to accumulate, at least when dealing with departments rather than individuals.

\section{Predictions for REF~2014}
\label{sectREF2014}

We next use the procedure described above to estimate group $h$-indices corresponding to REF~2014.
{Since these are based on the REF~2014 publication period, namely  from 2008 until the end of 2013, we employ the new notation $\hat{h}$ to distinguish the next results from the results for RAE~2008 data.}
We use the same list of higher education institutes as was used in RAE~2008, although that is sure to change to some degree for REF~2014.
Using the departmental $h$-indices as proxies, we build ranked lists of universities for the four  disciplines examined.
These are listed in  Tables~\ref{tab2}, \ref{tab2_Chem}, \ref{tab2_Phys} and \ref{tab2_Soc}.
It is interesting to compare the forecasted ranks with the previous ones based on RAE~2008, but already a weak correlation between $s$ and $h$-indices can cause the differences between two lists itself. Therefore, we compare rather the ranked lists of HEI's based on the $h_{2008}$ and on the {$\hat{h}_{2014}$} -- two indices, which were obtained by the same tools.
The arrows in the third columns ({$\hat{h}_{2014}$}) indicate the group $h$-index predictions for the direction of movement of the various HEI's at REF~2014 relative to their ranked positions based on their $h_{2008}$ values.

\begin{table}[!ht]
\caption{The list of British HEI's in \textbf{Biology}, ranked by RAE~2008-scores $s$, $h_{2008}$
and  by {$\hat{h}_{2014}$} {(the corresponding values of $h$-indices are shown in {parentheses})}. Scopus data were not available for some HEI's due to technical reasons and these are omitted from the corresponding lists. The ``up'' and ``down'' arrows show the direction of shift within the 3rd column relative to  the 2nd. The word `University' is omitted in the 2nd and the 3rd columns to save space.}
\begin{center}
\begin{tabular}{|l||l|l|}
\hline
\textbf{{HEI, ranked by }$s$}&\textbf{{HEI, ranked by }$h_{2008}$}&\textbf{{HEI, ranked by }$h_{2014}$}\\
\hline \hline
\parbox[t]{4.9cm}{
1. Institute of Cancer Research (ICR)\\
2. University of Manchester\\
3. University of Dundee\\
4. University of Sheffield\\
5. University of York\\
6. Imperial College London (ICL)\\
6. Kings College London (KCL)\\
8. Royal Holloway, \\$\hphantom{....}$University of London\\
9. University of Cambridge\\
10. University of Leeds\\
11. University of Edinburgh\\
11. University of Newcastle \\$\hphantom{......}$upon Tyne\\
13. Cardiff University\\
13. University of Aberdeen\\
13. University of Glasgow\\
16. University of St. Andrews\\
17. University of Bath\\
17. University of Birmingham\\
17. University of Durham\\
17. University of East Anglia\\
17. University of Exeter\\
17. University of Nottingham\\
23. University of Southampton\\
23. University of Warwick\\
25. University of Leicester\\
26. University of Liverpool\\
27. Queen Mary, University of London\\
27. University of Essex\\
29. University of Reading\\
29. University of Sussex\\
31. University of Kent\\
32. Queens University Belfast\\
33. Bangor University\\
34. University of Plymouth\\
35. University of Hull\\
36. Cranfield University\\
36. Swansea University\\
38. Liverpool John Moores University
}
&
\parbox[t]{3.2cm}{
1. Cambridge (111) \\
2. Edinburgh (91) \\
3. ICL (87) \\
4. KCL (86) \\
5. Dundee (84) \\
6. Glasgow (78) \\
7. ICR (73) \\
7. Birmingham (73) \\
9. Cardiff  (71) \\
10. Manchester (70) \\
11. Leicester (68) \\
12. Newcastle upon Tyne (66) \\
12. Sheffield (66) \\
14. Leeds (65) \\
15. York (62) \\
16. Southampton (61) \\
17. Nottingham (60) \\
18. Liverpool (57) \\
19. Sussex (53) \\
20. Bath (52) \\
20. Reading (52) \\
22. Aberdeen (50) \\
22. East Anglia (50) \\
24. Queens Belfast (47) \\
24. Warwick (47) \\
26. St. Andrews (45) \\
27. Durham (39) \\
28. Exeter (38) \\
29. Bangor  (37) \\
29. Queen Mary (37) \\
31. Royal Holloway (35) \\
32. Essex (34) \\
33. Hull (33) \\
34. Kent (31) \\
35. Cranfield  (29) \\
36. Swansea  (26) \\
36. Plymouth (26) \\
38. John Moores	(23)
}
&
\parbox[t]{3.3cm}{
1. Cambridge (143) \\
2. KCL $\uparrow$ (120) \\
3. ICL London (109) \\
4. Edinburgh $\downarrow$ (107) \\
5. Manchester $\uparrow$ (105) \\
6. Leeds $\uparrow$ (92) \\
7. Newcastle, Faculty \\$\hphantom{....}$of Medicine $\uparrow$ (89) \\
8. ICR $\downarrow$ (88) \\
9. Dundee $\downarrow$ (83) \\
10. Glasgow $\downarrow$ (82) \\
11. Birmingham $\downarrow$ (80) \\
12. Cardiff  $\downarrow$ (79) \\
12. Sheffield (79) \\
14.  Leicester $\downarrow$ (77) \\
14.  Nottingham $\uparrow$ (77) \\
16.  Southampton (72) \\
17. Liverpool $\uparrow$ (68) \\
18. Aberdeen $\uparrow$ (66) \\
19. York $\downarrow$ (64) \\
20. East Anglia $\uparrow$ (60) \\
21. Queens Belfast $\uparrow$ (59) \\
21. Exeter $\uparrow$ (59) \\
21.  Warwick $\uparrow$ (59) \\
24. Queen Mary $\uparrow$ (57) \\
25. St. Andrews $\uparrow$ (56) \\
26. Sussex $\downarrow$ (53) \\
27. Reading $\downarrow$ (52) \\
28. Bath $\downarrow$ (50) \\
29.Durham $\downarrow$ (46) \\
30. Bangor  $\downarrow$ (45) \\
31. Plymouth $\uparrow$ (41) \\
32. Swansea  $\uparrow$ (38) \\
32. Essex (38) \\
34. Royal Holloway $\downarrow$ (37) \\
35. Hull $\downarrow$ (36) \\
36. Cranfield  $\downarrow$ (34) \\
37.  John Moores  $\uparrow$ (32) \\
38. Kent $\downarrow$	(31)
}\\
\hline
\end{tabular} \label{tab2}
\end{center}
\end{table}

\begin{table}[!ht]
\caption{As in Table~2 but for \textbf{Chemistry}.
}
\begin{center}
\begin{tabular}{|l||l|l|}
\hline
\textbf{{HEI, ranked by }$s$}&\textbf{{HEI, ranked by }$h_{2008}$}&\textbf{{HEI, ranked by }$h_{2014}$}\\
\hline \hline
\parbox[t]{4.9cm}{
1. University of Cambridge\\
2. University of Nottingham\\
3. University of Oxford\\
4. University of Bristol\\
5. Imperial College London\\
6. University of Leeds\\
7. University of Liverpool\\
8. University of Manchester\\
8. University of York\\
8. University of Warwick\\
11. University of Durham\\
12. University of Sheffield\\
13. University College London\\
14. Cardiff University\\
15. University of Southampton\\
15. University of Birmingham\\
17. University of Bath\\
17. Heriot-Watt University\\
19. University of Sussex\\
20. University of East Anglia\\
21. Bangor University\\
22. University of Hull\\
23. Queens University Belfast\\
23. University of Newcastle \\$\hphantom{......}$upon Tyne\\
25. University of Leicester\\
25. University of Aberdeen\\
27. Loughborough University\\
28. University of Reading\\
29. University of Huddersfield
}
&
\parbox[t]{3.2cm}{
1. ICL (59) \\
2. Cambridge (58) \\
3. Oxford (54) \\
4. Bristol (48) \\
5. Manchester (45) \\
5. Nottingham (45) \\
5. Southampton (45) \\
8. Cardiff  (44) \\
8. UCL (44) \\
8. Liverpool (44) \\
11. Leeds (41) \\
12. Queens Belfast (40) \\
12. Sheffield (40) \\
14. Durham (39) \\
15. Warwick (38) \\
16. Birmingham (37) \\
16. York (37) \\
18. Sussex (36) \\
19. Hull (34) \\
20. Bath (33) \\
21. Reading (32) \\
22. Leicester (31) \\
22. Loughborough  (30) \\
24. Newcastle  (30) \\
25. East Anglia (29) \\
26. Aberdeen (24) \\
27. Heriot-Watt  (23) \\
28. Bangor  (17) \\
29. Huddersfield (15)
}
&
\parbox[t]{3.3cm}{
1. ICL (84) \\
1. Cambridge $\uparrow$ (84) \\
3. Oxford (74) \\
4. Manchester $\uparrow$ (66) \\
5. Liverpool $\uparrow$ (57) \\
6. UCL $\uparrow$ (55) \\
7. Bristol $\downarrow$ (52) \\
7. Leeds $\uparrow$ (52) \\
9. Durham $\uparrow$ (51) \\
9. Nottingham $\downarrow$ (51) \\
9. Warwick $\uparrow$ (51) \\
12. Southampton $\downarrow$ (50) \\
13. Cardiff  $\downarrow$ (49) \\
14. Bath $\uparrow$ (48) \\
15. York $\uparrow$ (46) \\
16. Birmingham (45) \\
17. Sheffield $\downarrow$ (44) \\
18. Queens  Belfast $\downarrow$ (43) \\
19. Newcastle  $\uparrow$\\
20. Heriot-Watt  $\uparrow$ (36) \\
21. Reading (35) \\
22. East Anglia $\uparrow$ (34) \\
23. Loughborough  $\downarrow$ (33) \\
24. Aberdeen $\uparrow$ (32) \\
25. Hull $\downarrow$ (29) \\
26. Leicester $\downarrow$ (26) \\
27. Sussex $\downarrow$ (25) \\
28. Bangor  (20) \\
28. Huddersfield $\uparrow$	(20)
}\\
\hline
\end{tabular} \label{tab2_Chem}
\end{center}
\end{table}

\begin{table}[!ht]
\caption{As in Table~2 but for  \textbf{Physics}}.
\begin{center}
\begin{tabular}{|l||l|l|}
\hline
\textbf{{HEI, ranked by }$s$}&\textbf{{HEI, ranked by }$h_{2008}$}&\textbf{{HEI, ranked by }$h_{2014}$}\\
\hline \hline
\parbox[t]{4.9cm}{
1. Lancaster University\\
2. University of Cambridge\\
2. University of Nottingham\\
2. University of St. Andrews\\
5. University of Bath\\
6. University of Edinburgh\\
7. Imperial College London (ICL)\\
8. University College London (UCL)\\
8. University of Durham\\
8. University of Glasgow\\
8. University of Sheffield\\
12. University of Birmingham\\
13. University of Oxford\\
13. University of Bristol\\
13. University of Liverpool\\
16. University of Manchester\\
16. University of Exeter\\
16. University of Sussex\\
19. University of Southampton\\
19. Heriot-Watt University\\
21. University of York\\
22. University of Warwick\\
22. University of Leicester\\
24. University of Leeds\\
25. Queen Mary, University of London\\
25. Loughborough University\\
27. University of Surrey\\
28. Swansea University\\
28. Kings College London (KCL)\\
30. Queens University Belfast\\
31. Cardiff University\\
32. University of Strathclyde
}
&
\parbox[t]{3.2cm}{
1. Cambridge (82) \\
2. ICL (80) \\
3. Oxford (70) \\
4. Birmingham (57) \\
5. UCL (54) \\
5. Bristol (54) \\
5. Durham (54) \\
5. Edinburgh (54) \\
5. Glasgow (54) \\
5. Manchester (54) \\
11. Queen Mary (53) \\
12. Liverpool (52) \\
13. Sussex (50) \\
14. Southampton (49) \\
15. St. Andrews (45) \\
16. Sheffield (44) \\
17. Lancaster  (40) \\
17. Queens  Belfast (40) \\
17. Leeds (40) \\
20. Bath (39) \\
21. KCL (37) \\
21. Surrey (37) \\
23. Nottingham (36) \\
24. Warwick (35) \\
25. Cardiff  (33) \\
26. Exeter (32) \\
27. Strathclyde (31) \\
28. Swansea  (30) \\
28. Leicester (30) \\
30. York (29) \\
31. Loughborough  (26) \\
32. Heriot-Watt  (25)
}
&
\parbox[t]{3.3cm}{
1. Cambridge (98) \\
2. Oxford $\uparrow$ (95) \\
3. ICL $\downarrow$ (94) \\
4. Manchester $\uparrow$ (84) \\
5. UCL (78) \\
6. Durham $\downarrow$ (73) \\
6. Edinburgh $\downarrow$ (73) \\
8. Glasgow $\downarrow$ (70) \\
9. Bristol $\downarrow$ (68) \\
9. Liverpool $\uparrow$ (68) \\
9. Southampton $\uparrow$ (68) \\
12. Birmingham $\downarrow$ (66) \\
13. Queen Mary$\downarrow$ (63) \\
14. St. Andrews $\uparrow$ (61) \\
15. Warwick $\uparrow$ (57) \\
16. Cardiff  $\uparrow$ (56) \\
16. Nottingham $\uparrow$ (56) \\
18. Sheffield $\downarrow$ (55) \\
19. Sussex $\downarrow$ (54) \\
20. Lancaster  $\downarrow$ (53) \\
21. Leeds $\downarrow$ (50) \\
21. Leicester $\uparrow$ (50) \\
23. Queens  Belfast $\downarrow$ (48) \\
24. Exeter $\uparrow$ (44) \\
24. Strathclyde $\uparrow$ (44) \\
26. Loughborough  $\uparrow$ (36) \\
26. Bath $\downarrow$ (36) \\
28. KCL $\downarrow$ (35) \\
29. Heriot-Watt  $\uparrow$ (34) \\
30. York (33) \\
31. Surrey $\downarrow$ (32) \\
32. Swansea  $\downarrow$ (30)
}\\
\hline
\end{tabular} \label{tab2_Phys}
\end{center}
\end{table}

\begin{table}[!ht]
\caption{As in Table~2 but for  \textbf{Sociology}.}
\begin{center}
\begin{tabular}{|l||l|l|}
\hline
\textbf{{HEI, ranked by }$s$}&\textbf{{HEI, ranked by }$h_{2008}$}&\textbf{{HEI, ranked by }$h_{2014}$}\\
\hline \hline
\parbox[t]{4.9cm}{
1. University of Manchester\\
2. University of Essex\\
2. Goldsmiths College, \\$\hphantom{....}$University of London\\
4. Lancaster University\\
5. University of York\\
6. University of Edinburgh\\
6. University of Surrey\\
8. University of Warwick\\
9. Cardiff University\\
9. University of Oxford\\
11. University of Sussex\\
12. University of Exeter\\
13. University of Cambridge\\
14. Open University\\
14. Queens University Belfast\\
16. Loughborough University\\
17. University of Aberdeen\\
18. London School of Economics \\$\hphantom{......}$and Political Science (LSEPS)\\
19. University of Newcastle \\$\hphantom{......}$upon Tyne\\
20. University of Nottingham\\
21. City University, London (CUL)\\
21. Brunel University\\
23. University of Bristol\\
24. University of East London\\
24. University of Glasgow\\
26. University of Leicester\\
27. University of Plymouth\\
27. Manchester Metropolitan \\$\hphantom{......}$University\\
29. Roehampton University\\
30. University of Liverpool\\
30. University of Birmingham\\
32. University of Teesside\\
33. University of Strathclyde\\
34. University of Huddersfield\\
35. University of the West of England, \\$\hphantom{......}$Bristol\\
35. Glasgow Caledonian University\\
37. Napier University\\
37. Robert Gordon University
}
&
\parbox[t]{3.2cm}{
1. Manchester  (32) \\
2. Oxford  (31) \\
3. Cardiff   (30) \\
4. Bristol  (29) \\
4. York  (29) \\
6. LSEPS (27) \\
6. Cambridge  (27) \\
8. Lancaster   (25) \\
8. Glasgow  (25) \\
10. Birmingham  (24) \\
10. Edinburgh  (24) \\
10. Newcastle  (24) \\
10. Nottingham  (24) \\
14. Warwick  (22) \\
15. Brunel   (21) \\
15. Open Uni.  (21) \\
15. Liverpool  (21) \\
18. Leicester  (20) \\
18. Surrey  (20) \\
20. CUL (19) \\
20. Aberdeen  (19) \\
20. Essex  (19) \\
20. Exeter  (19) \\
20. Sussex  (19) \\
25. Loughborough   (18) \\
26. Strathclyde  (17) \\
27. Plymouth  (16) \\
28. Queens  Belfast  (15) \\
29. Goldsmiths College  (14) \\
29. West of England, \\$\hphantom{......}$Bristol  (14) \\
31. Glasgow Caledonian   (12) \\
31. Manchester \\$\hphantom{......}$Metropolitan   (12) \\
31. Napier   (12) \\
34. East London  (11) \\
35. Robert Gordon   (9) \\
35. Huddersfield  (9) \\
35. Teesside  (9) \\
38. Roehampton (5)
}
&
\parbox[t]{3.3cm}{
1. Oxford $\uparrow$ (41) \\
2. Cambridge $\uparrow$ (36) \\
3. LSEPS $\uparrow$ (35) \\
3. Edinburgh $\uparrow$ (35) \\
3. Manchester $\downarrow$ (35) \\
6. Nottingham $\uparrow$ (34) \\
7. Cardiff  $\downarrow$ (31) \\
7. Sussex $\uparrow$ (31) \\
9. Lancaster  $\downarrow$ (30) \\
9. Birmingham $\uparrow$ (30) \\
9. Exeter $\uparrow$ (30) \\
12. Open Uni. $\uparrow$ (29) \\
12. York $\downarrow$ (29) \\
14. Bristol $\downarrow$ (28) \\
14. Glasgow $\downarrow$ (28) \\
16. Brunel  $\downarrow$ (27) \\
16. Newcastle $\downarrow$\\
16. Warwick $\downarrow$ (27) \\
19. Loughborough  $\uparrow$ (26) \\
19. Aberdeen $\uparrow$ (26) \\
21. Leicester $\downarrow$ (25) \\
21. Liverpool $\downarrow$ (25) \\
23. Surrey $\downarrow$ (24) \\
24. Essex $\downarrow$ (23) \\
24. Plymouth $\uparrow$ (23) \\
26. Strathclyde  (21) \\
27. Queens  Belfast $\uparrow$ (20) \\
28. West of England, \\$\hphantom{......}$Bristol  $\uparrow$ (19) \\
29. CUL $\downarrow$ (18) \\
30. Manchester \\$\hphantom{......}$Metropolitan  $\uparrow$ (17) \\
31. Glasgow Caledonian   (16) \\
32. Goldsmiths College $\downarrow$\\
32. Roehampton  $\uparrow$ (14) \\
34. Huddersfield $\uparrow$ (13) \\
35. Napier  $\downarrow$ (12) \\
35. East London $\downarrow$ (12) \\
35. Teesside  (12) \\
38. Robert Gordon  $\downarrow$ (9)
}\\
\hline
\end{tabular} \label{tab2_Soc}
\end{center}
\end{table}

\section{Conclusions}

There are persistent suggestions, primarily by research managers and policy makers, to replace or inform national peer-review research evaluation exercises by a simple system based on bibliometrics or scientometrics.
Such a set-up would have the advantage of being more cost effective and  less invasive.
However, to convince the academic research community of the reasonableness of such a system, it would need to have a proven high degree of accuracy relative to peer review because, besides its importance for funding purposes, such exercises -- and the inevitable rankings that follow them -- have predominant effects on institutional and departmental reputations.
One objection, frequently made about citation-based measures, is that they require a significant period of time to allow citations to accrue and thus every national evaluation would necessarily be ``historical''.

Here we have studied the correlations between two departmental quality metrics and the scores from RAE~2008. Of the two, the $h$- index performs better in terms of its similarity to that peer-review exercise. {At first sight, this is a surprising result} because the $h$-index is not an extensive or specific index (see, e.g., \cite{2008_Molinari}).
Moreover and also contrary to expectation, it is not required to wait a long time to  collect  sufficient numbers of citations -- the $h$-index calculated immediately after the specified publication period is as well correlated as that evaluated years later. {On the other hand, at least a part of the data which contribute to the immediate $h$-values has 7-years time spans, so many papers have past the peak of their citation record. This may account for the stability of the results.}

Based on these empirical findings, we then use the departmental $h$-index to make predictions for the rankings of universities in four different subject areas for REF~2014.
If the simple citation-based metric can, indeed, be used as some sort of proxy for the peer-review-based assessments, one would expect it to be able to predict the outcome, or some aspects of the outcomes, of REF~2014.
Even a limited degree of success may suggest that this metric could serve at least as a ``navigator'' for research institutes in between the massive expert exercises.

Here we delivered $h$-index predictions in advance of the outcomes of REF~2014. {We place the paper on the arXiv for the record and we will revisit it after the REF results are published to decide whether or not  there is any hope that a useful metric of this type could be developed, even as a ``navigator'' for managers and policy makers.}

\section*{Acknowledgements}
 This work was supported by the 7th FP, IRSES project No. 269139 ``Dynamics and cooperative phenomena in complex physical and biological environments'' and IRSES project No. 295302 ``Statistical physics in diverse realizations''.


\begin{thebibliography}{99}

\bibitem{Hirsch} Hirsch J.E., An index to quantify an individual's scientific research output. P. Nat. Acad. Sci. USA, 2005, \textbf{102}, No. 46, 16569--16572.


\bibitem{Dorothy_blog}
{Bishop~ D., BishopBlog. http://deevybee.blogspot.co.at/2013/01/an-alternative-to-ref2014.html
 Accessed 7 November 2014.}

\bibitem{Evidence_web}{Research Analytics,} 
http://www.evidence.co.uk. Accessed 31 October 2014.

\bibitem{MacRoberts1989}MacRoberts~M.H., MacRoberts~B.R., Problems of citation analysis: a critical review. J. Am. Soc. Inform. Sci., 1989, \textbf{40(5)}, 342--349.

\bibitem{Oppenheim1996}Oppenheim~Ch., Do citations count? Citation indexing and the
Research Assessment Exercise(RAE),  Serials: The Journal for the Serials Community, 1996, \textbf{9}, No. 2, 155--161.

\bibitem{Holmes2001}Holmes~A., Oppenheim~C., Use of citation analysis to predict the outcome of the 2001 Research Assessment Exercise for unit of assessment (UoA) 61: library and information management. Inform. Res., 2001, \textbf{6(2)}. Retrieved from http://informationr.net/ir/6-2/paper103.html  Accessed 31 October 2014.

\bibitem{2006_Raan}    Van Raan~A.F.J., Comparison of the Hirsch-index with standard bibliometric indicators and with peer judgment for 147 chemistry research groups, Scientometrics, 2006,  \textbf{67}, No.~3, 491--502.


 \bibitem{2013_Scientometrics} Mryglod O., Kenna R., Holovatch~Yu., Berche B., Absolute and specific measures of research group excellence, Scientometrics, 2013, \textbf{95(1)}, 115--127. DOI 10.1007/s11192-012-0874-7.

   \bibitem{2013_2_Scientometrics} Mryglod O., Kenna R., Holovatch~Yu., Berche B., Comparison of a citation-based
indicator and peer review for absolute and specific measures of research-group
excellence, Scientometrics, 2013, \textbf{97}, 767--777. DOI 10.1007/s11192-013-1058-9.

   \bibitem{RAE2008_web} RAE~2008: Research Assessment Exercise, http://www.rae.ac.uk/
Accessed 31 October 2014.

\bibitem{REF2014_web} REF~2014: Research Excellence Framework, http://www.ref.ac.uk/
Accessed 31 October 2014.

\bibitem{RAE_statement_panelE} RAE~2008. The panel criteria and working methods. Panel E. (2006). http://www.rae.ac.uk/pubs/2006/01/docs/eall.pdf. Accessed 31 October 2014.


\bibitem{HEFCE_web} Higher Education Funding Council for England: http://www.hefce.ac.uk/
Accessed 31 October 2014.

\bibitem{Enderby}
Enderby~J. Thick or thin? the funding dilemma. Editorial in the Journal of the Foundation for Science and Technology, 2011, \textbf{20}, No.~6, 3--4.

\bibitem{Evidence_2010}The future of the UK university research base. Evidence (a Thomson Reuters business) report, July 2010.


\bibitem{Evidence_2011}Funding research excellence: research group size, critical
mass \& performance. A University Alliance report, July 2011.

\bibitem{Schubert1996}Schubert A., Braun T., Cross-field normalization of scientometric indicators, Scientometrics, 1996, \textbf{36}, No.~3, 311--324.

\bibitem{Scopus_web} Scopus, http://www.scopus.com/. Accessed 31 October 2014.

\bibitem{Schreiber}Schreiber M., A variant of the h-index to measure recent performance, arXiv preprint arXiv:1409.3379 (2014).

\bibitem{2008_Molinari}Molinari~J.-F., Molinari~A., A new methodology for ranking scientific institutions, Scientometrics, 2008, \textbf{75}, No.~1, 163--174.



%
%
%
%
%
%
%
%
%
%



 \end{thebibliography}
\end{document}